\documentclass[12pt,a4paper,final]{iopart}

\usepackage{iopams}  
\usepackage{caption}
\usepackage{graphicx}
\usepackage[breaklinks=true,colorlinks=true,linkcolor=blue,urlcolor=blue,citecolor=blue]{hyperref}

\begin{document}

\title{Band Gap tunability in One-dimensional system}

\author{Payal Wadhwa}
\address{Department of Physics, Indian Institute of Technology Ropar, Nangal Road, Rupnagar, Punjab – 140001, India}

\author{Shailesh Kumar$^{1,2}$}
\address{$^1$Manufacturing Flagship, CSIRO, Lindfield West, New South Wales 2070, Australia}
\address{$^2$School of Chemistry, Physics and Mechanical Engineering, Queensland University of Technology, Brisbane, Queensland 4000, Australia}

\author{T.J. Dhilip Kumar}
\address{Department of Chemistry, Indian Institute of Technology Ropar, Rupnagar, Punjab - 140001, India}

\author{Alok Shukla$^{3,4}$}
\address{$^3$Department of Physics, Indian Institute of Technology Bombay, Powai, Mumbai - 400076, India}
\address{$^4$Department of Physics, Bennett University, Plot No 8-11, TechZone II, Greater Noida, Uttar Pradesh 201310, India}

\author[cor1]{Rakesh Kumar}
\address{Department of Physics, Indian Institute of Technology Ropar, Nangal Road, Rupnagar, Punjab – 140001, India}
\eads{\mailto{rakesh@iitrpr.ac.in}}

\begin{abstract}
The ability to tune the gaps of direct bandgap materials has tremendous potential for applications in the fields of LEDs and solar cells.\ However,\ lack of reproducibility of bandgaps due to quantum confinement observed in experiments on reduced dimensional materials,\ severely affects tunability of their bandgaps.\ In this letter,\ we report broad theoretical investigations of direct bandgap one-dimensional functionalized isomeric system using their periodic potential profile,\ where bandgap tunability is demonstrated simply by modifying the potential profile by changing the position of the functional group in a periodic supercell.\ It is verified for known synthetic,\ as well as natural polymers (biological and organic),\ and also for other one-dimensional direct bandgap systems.\ This insight would greatly help experimentalists in designing new isomeric systems of various bandgap values for polymers and one-dimensional inorganic systems for LEDs applications,\ and for effectively harvesting energy in solar cells.

\end{abstract}


\section{Introduction}

One-dimensional materials are in focus amongst the current research areas for their remarkable physical properties arising as a consequence of the reduced dimensionality.\ However,\ lack of control over reproducibility of bandgap values in one-dimensional materials \cite{PRL,Science,APL} is one of the challenges for its electronic applications like LEDs and LASER diodes,\ as it affects their bandgap tunability.\ Several theoretical studies have been reported to tune the bandgap of one-dimensional materials using various methods like strain,\ functionalization at the edges,\ doping,\ etc.\ \cite{Yan,Biel,Lu},\ however,\ the methods lack in control over the bandgap values.\ Since,\ bandgap is one of the most important factors while selecting a material for an electronic application;\ therefore,\ different direct bandgap materials have been explored for LEDs and LASER applications, e.g., Aluminium gallium nitride (AlGaN) for ultraviolet LEDs (below 400 nm),\ while Aluminium gallium arsenide (AlGaAs) for infrared LEDs (above 760 nm).\ Therefore,\ it would be of great interest if bandgap of a given material can be tuned,\ and this quest has been extended to polymers.\ Several experimental and theoretical studies on bandgap in polymers \cite{Patra,Hou,Bundgaard,W,Chem.Soc.Rev.,Chem.Mater1,Chem.Mater2} have been reported aimed at their applications in organic LEDs \cite{Hamid,Yu,AlSalhi} and solar cells \cite{Wang,Mori,Yiu,Woo,Cheng,Hou2,Bundgaard}.\ However,\ different values of bandgap noticed in experiments on isomeric polymers \cite{Nature,Huang,J.Phys.Chem.B} and also in theoretical studies \cite{Chem.Mater1,J.Phys.Chem.B,Bulletin of the Korean Chemical Society,Macromolecules} suggest that bandgaps may be tuned in one-dimensional system,\ if the underlying physics is understood.\ The fact that the work so far reported on bandgaps in isomeric polymers having the same functional group is still inconclusive \cite{J.Phys.Chem.B,Bulletin of the Korean Chemical Society,Macromolecules},\ motivated us to investigate and define the driving elements of different bandgaps seen in the isomeric systems.\ Once we understand the mechanism behind this, we may be in a position to tune the bandgaps of such materials as per our requirements.\\ 

In this work,\ investigations are carried out on one-dimensional isomeric polymers (synthetic and natural),\ and nanoribbons having the same functional groups.\ Since, polymers are large chain of monomers,\ therefore they are considered as one-dimensional periodic systems for band structure calculations \cite{Crit. Rev. Solid State Mater. Sci.,Synth.Met.,J.Chem.Educ.1,J.Chem.Educ.2,Int. J. Quantum Chem.}.\ Band structure calculations are performed using Density Functional Theory (DFT) as implemented in Vienna \textit{ab initio} simulation package (VASP) \cite{Comput.Mater.Sci.}.\ Generalized gradient approximation (GGA) \cite{GGA} is used for exchange-correlation of electron-electron interactions as implemented in projected augmented wave (PAW) formalism \cite{PAW}.\ Further,\ a vacuum layer of at least $\approx 15$ {\AA} is used to avoid interlayer interactions.\ The system is relaxed until a force on each atom in the unit cell is less than 0.001 eV.{\AA}$^{-1}$.\ k-mesh of size $25\times1\times1$ is used in Monkhorst-Pack formalism for momentum space sampling.

\section{Results and discussions}

\begin{figure}[h]
\centering
 \includegraphics[height=6cm]{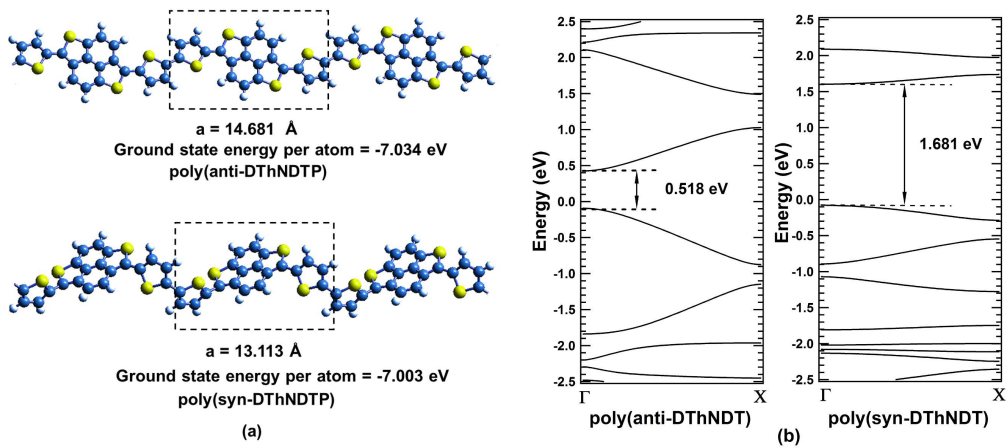}
  \caption{(color online) (a) Unit cells for poly(anti-DThNDT) and poly(syn-DThNDT) are represented in the dotted boxes.\ Blue,\ yellow,\ and white spheres represent carbon,\ sulfur,\ and hydrogen atoms,\ respectively.\ (b) Band structure plots corresponding to poly(anti-DThNDTP) and poly(syn-DThNDTP).}
  \label{fgr:str1}
\end{figure}

For investigating and defining the driving element of different bandgaps in isomeric systems,\ first example of low bandgap synthetic polymer polydithienyl naphthodithiophenes (DThNDT) $(C_{20}H_{8}S_{4})_{n}$ is considered.\ It exists in two isomeric forms,\ poly(anti-DThNDT) and poly(syn-DThNDT) \cite{J.Phys.Chem.B} having periodic unit cell of length 14.681 {\AA},\ and 13.113 {\AA},\ respectively [Figure \ref{fgr:str1} (a)].\ Cut-off energy of 500 eV is used for band structure calculations.\ Ground state energy calculated per atom for poly(anti-DThNDT) and poly(syn-DThNDT) are - 7.034 eV and - 7.003 eV,\ respectively,\ which are in close proximity being isomers.

Given the one-dimensional nature of polymers,\ their band structures are plotted from $\Gamma$ to X point [Figure \ref{fgr:str1} (b)].\ Direct bandgaps are observed at $\Gamma$ point for both the polymers of poly(anti-DThNDT) and poly(syn-DThNDT) with bandgaps of 0.518 eV and 1.681 eV, respectively.\ Bandgap for poly(anti-DThNDT) is smaller than that of poly(syn-DThNDT),\ which is in agreement with the experimental report \cite{J.Phys.Chem.B}.\ Band structures for these isomeric polymers are significantly different,\ even though they have same chemical formula,\ and practically the same ground state energy. 

\begin{figure}[h]
\centering
  \includegraphics[height=8cm]{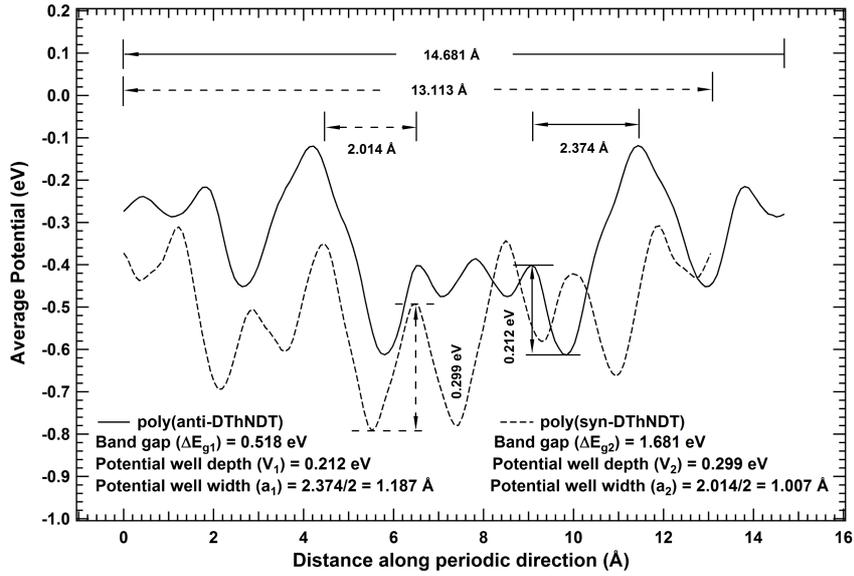}
  \caption{Average potential profile corresponding to the unit cell in the periodic direction of poly(anti-DThNDT) and poly(syn-DThNDT) are denoted in solid and dotted lines, respectively.}
  \label{fgr:str2}
\end{figure}

The polymers are one-dimensional systems with a repeating unit cell (monomer),\ therefore their bandgap may be related to their one-dimensional periodic potential profile similar to that of Kronig-Penney model.\ Since,\ potential is a scalar quantity,\ therefore average of potentials in the periodic direction of isomeric unit cell may be considered for comparative analysis of bandgap values.\ Average potential profile for poly(anti-DThNDT) and poly(syn-DThNDT) are plotted in the periodic direction as shown in Figure \ref{fgr:str2}.

The periodic average potential profiles of poly(anti-DThNDT) and poly(syn-DThNDT) are quite different to each other,\ and even to ideal rectangular potentials of Kronig-Penney model \cite{Proc.R.Soc.Lond.Math.Phys.Eng.Sci}.\ Since a system prefers to stay in its ground state;\ therefore the deepest potential well at global minimum in the periodic potential profiles are considered for comparative analysis of bandgap values \cite{Sci. Rep.} for isomeric systems.\ The global minimum for poly(anti-DThNDT) is located at 9.849 {\AA}, enclosed between two crests (barriers) located at 9.058 {\AA} and 11.432 {\AA}, while global minimum for poly(syn-DThNDT) is located at 5.517 {\AA},\ enclosed between two crests (barriers) located at 4.466 {\AA} and 6.480 {\AA}.\ From Figure \ref{fgr:str2},\ it can be seen that shape of the potential wells in the potential profile looks like inverse Gaussians,\ consisting of both well and barrier width.\ Therefore for simplifying the calculations,\ potential well is considered as square well potential of equal width for well and barrier (half of the distance between crests of the potential well).\ Since,\ depth of the deepest potential well at global minimum  {\lq{$V_{0}$}\rq},\ and its corresponding width {\lq{$a$}\rq} are finite and non-zero,\ distinct from the KP model (where $a \rightarrow 0$ and $V_{0}\rightarrow\infty$ for finite value of $V_{0}.a$) \cite{Proc.R.Soc.Lond.Math.Phys.Eng.Sci}.\ Hence,\ Schr\"odinger equation needs to be solved for the periodic square well potential of finite width and depth at global minimum to get first order bandgap,\ which may be correlated with the bandgap calculated using DFT.\ The energy eigenvalues corresponding to Schr\"odinger wave equation for an electron of mass {\lq{$m$}\rq} and energy {\lq{$E$}\rq} (where $E< V_{0}$) in a square well periodic potentials of finite depth {\lq{$V_{0}$}\rq} and width {\lq{$a$}\rq} can be obtained by solving the transcendental equation

\begin{equation} \label{eq:1}
\frac{\beta^{2}-\alpha^{2}}{2\alpha\beta}\sin(\alpha a)\sinh(\beta a)+\cos(\alpha a)\cosh(\beta a)=\cos(k.2a)\\
\end{equation}
where,
\begin{eqnarray}{\label{eqn:eqnad}}
\beta=\sqrt{\frac{2mV_{0}}{\hbar^{2}}-\alpha^{2}}, 
\frac{\hbar^{2}\alpha^{2}}{2m}=E+V_{0}>0
\end{eqnarray}
and, 
\begin{eqnarray}{\label{eqn:eqnad}}
\frac{\hbar^{2}\beta^{2}}{2m}=-E<0
\end{eqnarray}

For isomeric polymers of polydithienyl naphthodithiophenes (DThNDT) $(C_{20}H_{8}S_{4})_{n}$, width and depth of the deepest potential well at global minimum for poly(anti-DThNDT) are 1.187 {\AA} and 0.212 eV,\ respectively,\ while for poly(syn-DThNDT) are 1.007 {\AA} and 0.299 eV,\ respectively (Figure \ref{fgr:str2}).\ Using these values in equation \ref{eq:1},\ it is found that poly(syn-DThNDT) has larger bandgap than that of poly(anti-DThNDT),\ which is in agreement with bandgap values calculated using DFT,\ and experimental reports \cite{J.Phys.Chem.B}.\ Therefore,\ it is concluded that bandgap of one-dimensional isomeric systems may be correlated with depth and width of the potential well at global minimum in the periodic average potential profile.   

\begin{figure}[h]
\centering
  \includegraphics[height=8cm]{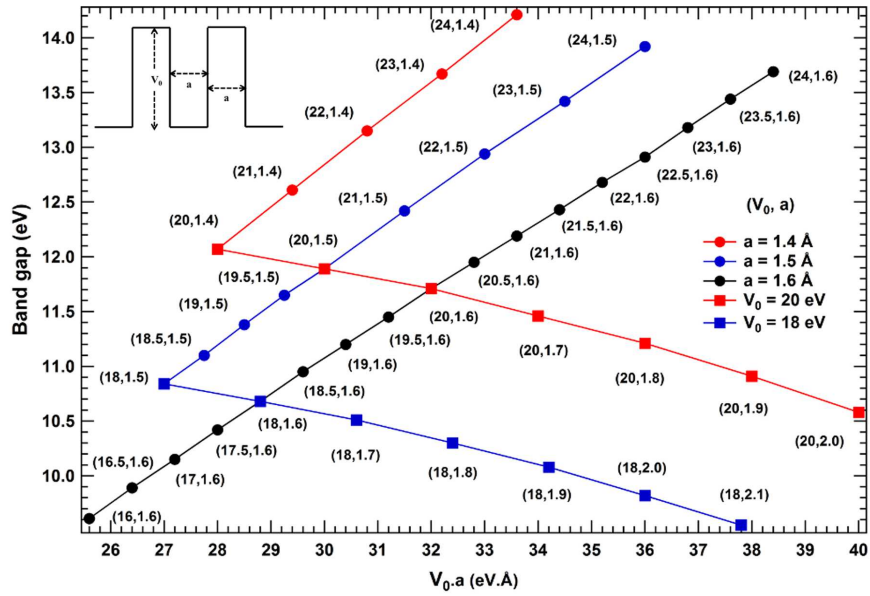}
  \caption{(color online) First order bandgap plot as a function of {\lq{$V_{0}.a$}\rq} for different values of {\lq{$V_{0}$}\rq} and {\lq{$a$}\rq} for the bound systems corresponding to a square well potential represented in the inset.}
  \label{fgr:str3}
\end{figure}

In order to find out how bandgap of one-dimensional isomeric systems may be correlated with their deepest potential well at global minimum in the periodic direction;\ a general correlation needs to be formulated and establish its validity for other isomeric systems.\ Since,\ isomeric systems usually would have different dimensions ($V_{0}$ and $a$) of the deepest potential well at global minimum,\ therefore the transcendental equation \ref{eq:1} needs to be solved for bound states of different {\lq{$V_{0}.a$}\rq} varying both {\lq{$V_{0}$}\rq} and {\lq{$a$}\rq}.\ In fact for bound states ($E< V_{0}$),\ depth of potential well {\lq{$V_{0}$}\rq} should be greater than a critical value for a given width {\lq{$a$}\rq},\ which we have considered in the calculations;\ for example,\ the critical value of depth calculated for potential well is 19.04 eV for a width of 1.4 {\AA},\ while critical depth turns out to be 16.59 eV for a width of 1.5 {\AA}.\ Bandgap for bound state periodic square well potential obtained from energy-momentum dispersion relations on solving the transcendental equation \ref{eq:1},\ for different {\lq{$V_{0}.a$}\rq} are plotted in Figure \ref{fgr:str3}.

On comparing the bandgap values of two isomeric systems with their corresponding width ({\lq{$a$}\rq}) and depth ({\lq{$V_{0}$}\rq}) of potential well e.g.,\ ($\triangle E_{g1}, V_{1},a_{1}$)\ and ($\triangle E_{g2}, V_{2},a_{2}$) for different product of {\lq{$V_{0}.a$}\rq}, following correlations are derived \\
\textbf{Case I} $(a_{1}=a_{2},V_{1}=V_{2})$ \\
then  $\triangle E_{g1}=\triangle E_{g2}$ \\
\textbf{Case II}  $(a_{1}=a_{2},V_{1}>V_{2})$ \\
then  $\triangle E_{g1}>\triangle E_{g2}$ \\
\textbf{Case III}  $(a_{1}>a_{2},V_{1}=V_{2})$ \\
then  $\triangle E_{g2}>\triangle E_{g1}$ \\
\textbf{Case IV}  $(a_{1}>a_{2},V_{2}>V_{1})$ \\
then  $\triangle E_{g2}>\triangle E_{g1}$ \\
\textbf{Case V}  $(a_{1}>a_{2},V_{1}>V_{2})$ \\
In this case,\ sign of slopes for bandgap as a function of {\lq{$V_{0}.a$}\rq} may change on changing {\lq{$V_{0}$}\rq} and {\lq{$a$}\rq} w.r.t.\ a reference point,\ therefore bandgap correlation can be predicted only on solving equation \ref{eq:1} for corresponding {\lq{$V_{0}$}\rq} and {\lq{$a$}\rq}. 

For isomeric polymers of polydithienyl naphthodithiophenes (DThNDT) $(C_{20}H_{8}S_{4})_{n}$,\ the width and depth of the deepest potential well at global minimum in the periodic potential profile for poly(anti-DThNDT) are 1.187 {\AA} (say {\lq{$a_{1}$}\rq}) and 0.212 eV (say $V_{1}$),\ respectively,\ while for poly(syn-DThNDT) are 1.007 {\AA} (say {\lq{$a_{2}$}\rq}) and 0.299 eV (say $V_{2}$),\ respectively [Figure \ref{fgr:str2}].\ Since $a_{1} > a_{2}$ and $V_{2} > V_{1}$,\ therefore according to correlations of {\bf Case IV},\ poly(syn-DThNDT) should have larger bandgap than poly(anti-DThNDT);\ which agrees with our band structure calculations using DFT,\ and other experimental reports \cite{J.Phys.Chem.B}.\ The agreement of derived correlation with theoretical and experimental results establishes its validity.\\ 

To verify it further,\ the investigation is extended to other isomeric synthetic polymer polydialkylterthiophenes $(C_{36}H_{54}S_{3})_{n}$ [see supplementary material 1] and  natural polymers (biopolymers and organic polymers) [see supplementary material 2 and 3],\ bandgaps are found to be correlated with the dimension of the potential well at global minimum as per the derived correlations.\ The investigations have been extended to natural polymers for extensive validity of the correlations,\ even though they are insulating and of little importance to electronic applications. 

On the basis of theoretical analyses,\ it is established that bandgap of isomeric systems are correlated with width and depth of the deepest potential well at global minimum in their periodic potential profile.\ From derived correlations,\ it may be predicted that bandgap of one-dimensional periodic system may be tuned,\ if width and depth of the deepest potential well in the periodic potential profile is altered on changing the position of functional group in the periodic unit cell. 

\begin{figure}[h]
\centering
  \includegraphics[height=5cm]{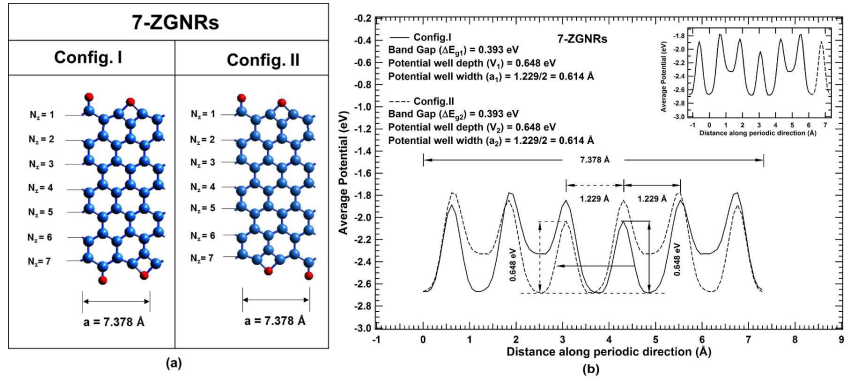}
  \caption{(color online) (a) Unit cells corresponding to two configurations Config.\ I and Config.\ II of 7-ZGNRs, where blue and red spheres represent carbon and oxygen atoms,\ respectively.\ (b) Average potential profile of the unit cell for Config.\ I and Config.\ II in the periodic direction of 7-ZGNRs are denoted with solid and dotted curve,\ respectively.\ Inset shows the overlap of potential profiles on relative shifting along the periodic direction.}
  \label{fgr:str4}
\end{figure}

To establish the concept of bandgap tunability in one-dimensional systems;\ the investigation is extended to theoretical GNRs (same molecular formula for the unit cells) having the same functional group in the periodic unit cell but of different arrangements. Zigzag GNRs (ZGNRs) of the same width functionalized at the edges with oxygen atoms in two typical ways [say Config.\ I and Config.\ II as shown in Figure \ref{fgr:str4}(a)] are considered for calculations. {\it$sp^{2}$} and {\it$sp^{3}$} hybridized carbon atoms are considered at edges for visible distinction of isomeric change in the periodic unit cell (7.378 {\AA}) of zigzag GNRs. Typical edge configurations of ZGNRs ($N_{z}$ = 7) are shown in Figure \ref{fgr:str4}(a).\ Periodic average potential profiles corresponding to Config.\ I and Config.\ II are plotted in Figure \ref{fgr:str4}(b).\ Even though their average potential profiles look different,\ their potential profiles superpose on each other on relative shifting in the periodic direction.\ The width and depth of potential well at global minimum for Config.\ I and Config.\ II are exactly same 0.614 {\AA} ($a_{1} = a_{2}$) and 0.648 eV ($V_{1} = V_{2}$).\ Since $a_{1} = a_{2}$ and $V_{1} = V_{2}$,\ therefore according to the derived correlations ({\bf Case I}),\ bandgap of both the configurations of ZGNRs should be equal.
To check the validity of the correlations,\ band structure calculations for 7-ZGNRs are performed with cut-off energy of 450 eV.\ Band structures are plotted from $\Gamma$ to X point (Figure 5) for $N_{z}$ = 7.\ Direct bandgap of 0.393 eV is observed at $\Gamma$ point for both the configurations.\ It is found that bandgap values of ZGNRs for Config.\ I and Config.\ II are same,\ even ground state energy per atom of the unit cell calculated for both the configurations are same,\ - 8.762 eV.\ On further analyses of potential profiles of ZGNRs,\ the same correlations are found to be hold for other odd ZGNRs ($N_{z}$ = 3 to 17),\ which are verified with the band structure calculations using DFT. 

\begin{figure}[h]
\centering
  \includegraphics[height=6cm]{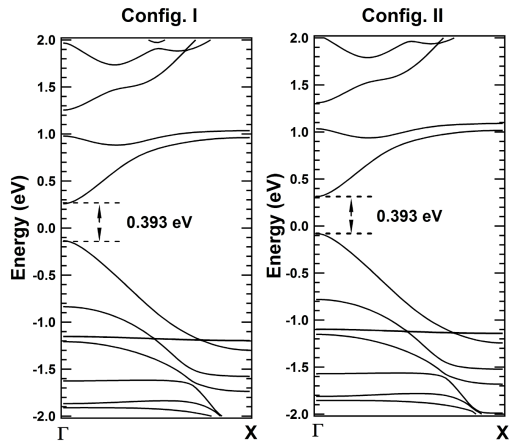}
  \caption{Band structure plots corresponding to Config.\ I  and  Config.\ II.}
  \label{fgr:str5}
\end{figure}
However,\ potential profiles of even $N_{z}$-ZGNRs are different for both the configurations of the same functional groups,\ therefore bandgaps are expected to be different according to the derived correlations;\ and is verified from band structure calculations using DFT [See supplementary material 4 and 5],\ which justifies tunability of bandgap value in direct bandgap one-dimensional systems.

\section{Conclusions}

On the basis of theoretical analyses of one-dimensional having the same functional group in the periodic unit cell,\ but of different arrangements,\ it is observed that
\begin{itemize}
  \item Bandgap of one-dimensional systems are correlated to the depth and width of potential well at global minimum in the periodic potential profile.
  \item The correlations derived between bandgap and dimension of periodic potential well at global minimum is verified for known isomeric systems of synthetic as well as natural polymers (biological and organic),\ and bandgap tunability is also established for one-dimensional nanoribbons.
\end{itemize}
Finally,\ it is concluded that bandgap of one-dimensional system can be tuned by changing the position of functional group in the periodic unit cell of the same material,\ which may be used for designing materials of different bandgap values for LEDs applications and effectively harvesting energy in solar cells;\ and insight may be extended to understand the different physical properties of isomers of biopolymers such as proteins.

\section*{Acknowledgements}

The author thanks IIT Ropar for providing the High Performance Supercomputing facility.\ We also thank S.\ Kaur for doing some preliminary calculations related to manuscript.

\end{document}